\documentclass[preprint,prc,aps]{revtex4}
\usepackage{latexsym}
\everymath={\displaystyle}
\usepackage{rotating}
\usepackage{young}
\def\1{\mbox{l\hspace{-0.53em}1}}

\begin{document}
\title{Group theoretical study of 
nonstrange and strange  
mixed symmetric baryon states $[N_c-1,1]$ in the $1/N_c$ expansion}

\author{N. Matagne$^a$\footnote{e-mail address: nicolas.matagne@umons.ac.be}}

\author{Fl. Stancu$^b$\footnote{e-mail address: fstancu@ulg.ac.be
}}
\affiliation{
$^a$ University of Mons, Service de Physique Nucl\'eaire et Subnucl\'eaire,Place du Parc 20, B-7000 Mons, Belgium \\
$^b$ University of Li\`ege, Institute of Physics B5, Sart Tilman,
B-4000 Li\`ege 1, Belgium}

\date{\today}

\begin{abstract}
\baselineskip=0.50cm
Using group theory arguments we extend and complete our previous work
by deriving all SU(6) exact wave functions associated to the spectrum of   
mixed symmetric baryon states $[N_c-1,1]$ in the $1/N_c$ expansion.
The extension to SU(6) enables us to study   the mass spectra of both
strange and nonstrange baryons,
while previous work was restricted to nonstrange baryons described by
SU(4).
The wave functions are specially written in a form to allow a comparison 
with the approximate, 
customarily used wave functions, where
the system is separated into a ground state core and an excited quark. 
We show that the matrix elements of the flavor operator calculated
with the exact wave functions acquire the same asymptotic form at 
large $N_c$, irrespective of the spin-flavor multiplet contained in
$[N_c-1,1]$, while with the approximate wave function one cannot obtain
a similar behaviour.  
The isoscalar factors of the
permutation group of $N_c$ particles derived here can be used in any
problem where a given fermion system is described by the partition $[N_c-1,1]$,
and one fermion has to be separated from the rest.

\end{abstract}

\maketitle

\section{Introduction}
The $1/N_c$ expansion method \cite{HOOFT,WITTEN,Gervais:1983wq,DM},
where $N_c$ is the number of colors, has lead 
to a better understanding of the spin-flavor structure of baryons
in the context of QCD. Much  work has been devoted to the ground  
state of light \cite{DJM94,DJM95,CGO94,Jenk1,JL95,DDJM96} and heavy baryons
\cite{Jenkins:1996de,Jenkins:2007dm}. For $N_c \rightarrow \infty$
the ground state is governed by an exact contracted 
SU(2$N_f$) symmetry \cite{Gervais:1983wq,DM},
where $N_f$ is the number of flavors. Accordingly, baryon masses 
are degenerate at $N_c \rightarrow \infty$. For finite $N_c$ the mass splitting
starts at order  $1/N_c$. It is customary to drop higher order corrections
in the mass formula. The results on spectra and mass relations prove that 
the large $N_c$ world is sufficiently close to $N_c$ = 3.

For ground state baryons the study is systematic and 
straightforward because the orbital wave function is  
symmetric and also irrelevant in the calculations. The spin-flavor wave 
function is also symmetric which makes quite easy to deal with it.

For excited states the symmetry is enlarged to 
SO(3) $\times$ SU(2$N_f$) to include orbital excitations. Then the system 
acquires a new 
degree of freedom, described by a specific orbital wave function,
the symmetry of which must match that of the spin-flavor  part,
in order to lead to a totally symmetric state in the orbital-spin-flavor space. 
The excited states described by symmetric wave functions in both the
orbital and spin-flavor degrees of freedom are nearly as simple 
as the ground state. Results are available for
the Roper resonance 
$[\textbf{56'},0^+]$   \cite{CC00}, and for states belonging to the  
$[\textbf{56},2^+]$ \cite{GSS03}
and to the $[\textbf{56},4^+]$ \cite{MS1} multiplets respectivelys.
For finite $N_c$ the mass splitting starts at order  $1/N_c$, like for
the ground state. 

More complicated are the mixed symmetric states in both orbital and
spin-flavor space. They belong to the  $[\textbf{70},\ell^P]$ multiplet
with parity $P = (-)^{\ell}$. Starting from group 
theoretical arguments, here we study the SU(6)  $[N_c-1,1]$ 
multiplet for arbitrary $N_c$ and thus complete 
our previous work \cite{Matagne:2008fw}, restricted to SU(4)
(for a review see 
\cite{Matagne:2008zn} and \cite{Matagne:2009hd}).
We recall that the SU(6) generators are $S^i$, $T^a$ and $G^{ia}$
acting on spin, flavor and spin-flavor respectively.
   
So far the most extensively studied  
is the $[{\bf 70},1^-]$ multiplets.  
This is the simplest and the best known experimentally group of
states. Historically, the first 
approach, presently a standard procedure, was based on the decoupling of 
the system into a ground state core described by a symmetric spin-flavor 
state of $N_c - 1$ quarks and an orbitally excited quark
\cite{Goi97,PY,CCGL,CGKM,CaCa98,Goity:2002pu}.
Accordingly  the SU(2$N_f$) generators were written as 
\begin{equation}\label{generators}
S^i = S^i_c + s^i, ~~T^a = T^a_c + t^a, ~~G^{ia} = G^{ia}_c + g^{ia},
\end{equation}
($i = 1,2,3$ and $a = 1,2,\ldots,8$),
where the operators carrying the lower index $c$ act on the core and 
the lower case operators act on the excited quark. 
This method is 
in the spirit of the Hartree picture and 
the system is described by an approximate wave function, where the orbital
part has a configuration of type $s^{N_c-1} p$ (no antisymmetrization) 
which is combined with an approximate (truncated) spin-flavor part. 
The splitting (\ref{generators}) of the generators amounts to an excessively 
large number of independent operators to be included in the mass formula,
difficult to handle when the data is restricted, as it usually happens.  
As being the first proposal in large $N_c$ baryon spectroscopy, we have also 
applied it to the
$[\textbf{70},\ell^+]$ multiplet ($\ell$ = 0, 2) 
\cite{Matagne:2005plb,Matagne:2006zf}. Consistenly with previous studies,
we found that the splitting starts at order
$\mathcal{O}(N^0_c)$. There are many interesting papers
in the field to be cited. However here we have to restrict the list to our
specific goal. 

From the studies we perfomed on the $[\textbf{70},\ell^+]$ multiplet, 
we understood that a simpler procedure can as correctly be used, where 
no quark is decoupled from the system, all identical quarks being treated on 
the same footing, with an exact wave function in the orbital-flavor-spin
space \cite{Matagne:2006dj}. We found out that the key to the problem was 
the knowledge of the matrix elements of all SU(2$N_f$) generators, $S^i$, 
$T^a$ and $G^{ia}$ for mixed symmetric states $[N_c-1,1]$. For SU(4) they 
were derived in Ref. \cite{Matagne:2006dj} and for SU(6) in 
Ref. \cite{Matagne:2008kb}.
In the standard procedure the problem was simplified, 
by truncating the wave function to a part where 
a quark is decoupled from the whole system, the rest
remaining in a ground state symmetric core. In this way the problem 
was reduced to the knowledge of the matrix elements of SU(2$N_f$) 
generators for the core, in the symmetric representation $[N_c-1]$. 

To better clarify our purpose, let us give an example in the SU(4)
standard procedure, in connection to the isospin operator.
The SU(2)-isospin Casimir operator was written as
$T^2 = T^2_c + 2 t \cdot T_c + 3/4$, \emph{i.e.} formed of 
three independent pieces, corresponding to the terms in this decomposition. 
In SU(4) $T^2_c$ and $S^2_c$ have identical matrix elements because
the spin and isospin states of a symmetric core are identical, so that
$T^2_c$ can be neglected. But $t \cdot T_c$ has different matrix
elements from $s \cdot S_c$ as one can clearly see from Table II
of Ref. \cite{CCGL}. Then, in the decoupling scheme
the isospin can be introduced only through $t \cdot T_c$.
In Ref. \cite{Matagne:2008fw} Table VI
we have shown that the introduction of the operators 
$\frac{1}{N_c}t \cdot T_c$ together with  $\frac{1}{N_c}S^2_c$ 
and $\frac{1}{N_c}s \cdot S_c$ separately deteriorates the fit.
This may explain why $\frac{1}{N_c}t \cdot T_c$ has been avoided in previous 
numerical fits in SU(4) \cite{CCGL}.

Physical consequences of the neglect  of the isospin operator 
were discussed in Ref. \cite{Matagne:2006dj} for SU(4), where it was
shown that the isospin term, neglected in the standard procedure,
becomes as dominant in $\Delta$ as the  spin term  in $N$ resonances.
As a first physical application of this work, where we extend our 
procedure to SU(6) we ask again the question why the operator 
$\frac{1}{N_c}t \cdot T_c$, as well as $T^2_c$ were avoided in previous
numerical fits  in SU(6) \cite{Goity:2002pu}.

Before presenting our work we wish to point out that the calculation
of the matrix elements of the operators appearing in the mass formula, with
the approximate wave function of the standard procedure
is not however an approximation. In the framework
of a large $N_c$ quark model, by using properties of the permutation
group, Pirjol and Schat \cite{Pirjol:2007ed} have shown that on can pass 
from the exact wave function to that of Ref. \cite{CCGL}  
without making any approximation.
This implies that an approximate wave function can be used in an effective 
theory, provided the constraints imposed by a given quark model are
satisfied. These constraints represent fixed ratios between specific 
coefficients in the mass formula in terms of well defined radial integrals. 

By analogy to our previous work \cite{Matagne:2008fw}, 
here we analyze the spin and 
flavor terms both for the exact  and the truncated, or else, approximate, wave
function of the $[\textbf{70},\ell^P]$ multiplet, 
without any prejudice. In other words we extend  
our previous work from SU(4) ($N_f$ = 2) to SU(6)~ ($N_f$ = 3). 
We briefly recall the standard procedure based on the core+excited quark 
separation. The relation between the approximate  
wave function \cite{CCGL} and the exact one, 
has been presented already in Ref.\ \cite{Matagne:2008fw}.
To test the validity of the approximate wave function 
we compare the analytical expressions of the matrix elements
of the spin and flavor operators entering the mass formula. For the 
first time we explicitly show that at large $N_c$
the approximate wave function provides matrix elements of the flavor operator 
where the flavor 
singlet $^21$  behaves asymtotically different  from  the octets $^28$, 
$^48$ and the decuplet $^210$ (the notation corresponding to $N_c$ = 3), 
which means that the large $N_c$ counting rule is broken
(see Sec. \ref{matelem}).
By contrast, the exact wave functions lead to identical analytic forms at
large $N_c$  for all spin-flavor multiplets belonging to the  representation
$\bf 70$ of SU(6) allowing a consistent definition of the flavor operator.

The paper is organized as follows. In the next section we recall the 
relation between the exact and approximate wave function and derive
the isoscalar factors needed for the flavor singlet $^21$. 
In Sec. III we derive analytic expressions of some dominant operators
entering the mass formula and discuss their behavior at large $N_c$
for the multiplets which become $^28$,  $^48$,  $^210$ and $^21$
when $N_c$ = 3, both for the approximate and the exact wave function.
The last section presents our conclusions.
In Appendix A we recall some isoscalar factors obtained previously, but
needed in this work. In Appendix B we describe the procedure
to obtain general analytic expression of  new S$_N$ isoscalar factors 
associated with the wave function of the $^21$ multiplet.
Appendix C is devoted to the derivation of the matrix elements
of the generator $G^{ja}$  of SU(6) and exhibits the SU(6) isoscalars
factors calculated in this work. 

\section{The wave function}\label{WF}

We deal with a system of $N_c$ quarks where one quark carries $\ell$ units of
orbital excitation. Therefore the orbital ($O$) wave function  must have 
a mixed symmetry $[N_c-1,1]$, which describes the lowest excitations in
a baryon.
The $N_c-1$ independent basis states of the $[N_c-1,1]$ irreducible 
representation (irrep)  corresponding to  $N_c-1$ 
Young tableaux, as presented below, is equivalent to a basis
written in terms of $N_c-1$  internal Jacobi coordinates, thus
the center of mass motion is automatically removed.
The center of mass motion is then described by the symmetric state $[N_c]$
with one excited quark.

The color wave function being antisymmetric, the
orbital-spin-flavor wave part must be symmetric. Then the spin-flavor ($FS$)
part must have the same symmetry as the orbital part in order to obtain a 
totally symmetric
state in the orbital-spin-flavor space. We recall that
the general form of such a wave function is \cite{Stancu:1991rc}
\begin{equation}
\label{EWF}
|[N_c] \rangle = {\frac{1}{\sqrt{d_{[N_c-1,1]}}}}
\sum_{Y} |[N_c-1,1] Y \rangle_{O}  |[N_c-1,1] Y \rangle_{FS},
\end{equation}    
where $d_{[N_c-1,1]} = N_c - 1$ is the dimension of the representation 
$[N_c-1,1]$ of the permutation group $S_{N_c}$ and $Y$ is a symbol for a
Young tableau (Yamanouchi symbol). 
The sum is performed over all possible standard Young tableaux. 
By convention, in each term the first basis vector represents the orbital  
space  and the second the spin-flavor space.
In this sum there is only one $Y$ (the normal Young tableau) 
where the last particle is in 
the second row and $N_c - 2 $ terms where the last particle
is in the first row.
The explicit form of the orbital part is not needed in this work.

More precisely, we write $Y = (pqy)$ where $p$ is the row of
the $N_c$-th particle, $q$ the row of the  $(N_c-1)$-th particle and $y$
is the Young tableau of the remaining particles. 
Let us denote by $p$, $p'$ and $p''$ the position of
the last particle in the spin-flavor, spin and flavor Young tableaux
respectively. They are indicated by crosses in the example 
given by Eqs. (\ref{singletch52})-(\ref{prime}) below.
Similarly for the $(N_c-1)$-th particle we have $q$, $q'$ and $q''$ and for the
rest $y$, $y'$ and $y''$.

We need now to decompose the spin-flavor wave function into its spin and flavor
parts. 
For this purpose we use the Clebsch-Gordan (CG) coefficients of $S_{N_c}$, 
denoted by $S([f']p'q'y' [f'']p''q''y'' | [f]pqy )$ 
and their factorization property \cite{Stancu:1991rc}.    
Denoting by $K([f']p'[f'']p''|[f]p)$ the isoscalar factors of $S_{N_c}$ 
we have \cite{Matagne:2008fw}
\begin{equation}\label{racah} 
S([f']p'q'y' [f'']p''q''y'' | [f]pqy ) =
K([f']p'[f'']p''|[f]p) S([f'_{p'}]q'y' [f''_{p''}]q''y'' | [f_p]qy ),
\end{equation}
where the second factor in the right-hand side is a CG coefficient of
$S_{N_c-1}$ containing the 
partitions $[f'_{p'}]$, $[f''_{p''}]$ and $[f_p]$ obtained after 
the removal of the $N_c$-th quark.
Keeping in mind that, for a given  $p$, the quantum numbers of the SU(6)
wave function are the same
and by using the above property we can write the spin-flavor part of the wave 
function as
\begin{eqnarray}\label{fs}
|[N_c-1,1]p;(\lambda\mu) Y I I_3;S S_3 \rangle = 
\sum_{p' p''} K([f']p'[f'']p''|[N_c-1,1]p) 
|S S_3;p' \rangle|(\lambda\mu) Y I I_3;p'' \rangle,  
\end{eqnarray}
where 
$|S S_3;p' \rangle|(\lambda\mu) Y I I_3;p'' \rangle$
contains the CG coefficients $S([f'_{p'}]q'y' [f''_{p''}]q''y'' | [f_p]qy )$
and includes a sum over $q' y'$ and $q'' y''$. These CG coefficients
sum up to 1 by normalization. Then in the matrix elements of every 
SU(6) opearator we shall have one
term with $p = 2$ and $N_c - 2$ terms with $p = 1$ (see example in the next 
section).

In the wave function (\ref{fs})
the spin part $|S S_3; p' \rangle$ is defined by the SU(2) coupling 
\begin{equation}\label{spin}
|S S_3; p' \rangle = \sum_{m_1,m_2}
 \left(\begin{array}{cc|c}
	S_c    &    \frac{1}{2}   & S   \\
	m_1  &         m_2        & S_3
      \end{array}\right)
      |S_cm_1 \rangle |1/2m_2 \rangle,
\end{equation}
with $S_c = S - 1/2$ for $p' = 1$    and $S_c = S + 1/2$ for   $p' = 2$
and the flavor part by the SU(3) coupling 
\begin{eqnarray}\label{flavor}
\lefteqn{|(\lambda\mu) Y I I_3,p'' \rangle =} \nonumber \\ & &
\sum_{Y_c,I_c,I_{c_3},\atop y,i,i_3} \left(\begin{array}{cc||c}
	(\lambda_c\mu_c)    &    (10)   &  (\lambda\mu)  \\
	Y_cI_c &         yi        & YI
      \end{array}\right)
\left(\begin{array}{cc|c}
	I_c    &    i   & I   \\
	I_{c_3}  &   i_3              & I_3
      \end{array}\right)
|(\lambda_c\mu_c)Y_cI_cI_{c_3}\rangle |(10)yii_3 \rangle,
\end{eqnarray}
with $(\lambda_c,\mu_c)$  = $(\lambda-1,\mu)$ for $p''$ = 1,
 $(\lambda_c,\mu_c)$  = $(\lambda+1,\mu-1)$ for $p''$ = 2 and  
 $(\lambda_c,\mu_c)$ = $(\lambda,\mu+1)$ for $p''$ = 3.
Here $\lambda$ and $\mu$ are consistent with the partition
$[f'']$ from Tables  \ref{spinonehalf}, \ref{spin1/2n}, \ref{spin3/2n}
and \ref{spin1/2delta} respectively.
As usually, in Eq. (\ref{flavor}), the SU(3) CG coefficient has been factorized 
into an isoscalar factor and an SU(2)-isospin factor \cite{DESWART}.

By taking $N_c = 7$ let us first illustrate Eq. (\ref{fs}) in terms of Young  
tableaux for the case presently under study, namely the flavor singlet of 
spin $S$ = 1/2.  We have two linearly independent spin-flavor 
states 
\begin{eqnarray}
\label{singletch52}
\raisebox{-9.0pt}{\mbox{\begin{Young}
 & & & & & \cr
$\times$ \cr
\end{Young}}}\
&=& K([43]1[331]3|[61]2)\! \! \!
\raisebox{-9.0pt}{\mbox{
\begin{Young}
& & & $\times$\cr
& & \cr
\end{Young}}} \ \times \! \! \! \! \!
\raisebox{-15pt}{\mbox{
\begin{Young}
& & \cr
& & \cr
$\times$ \cr
\end{Young}}}~ ,\ 
\\
\nonumber \\
\vspace{0.5cm}\label{prime}
\raisebox{-9.0pt}{\mbox{\begin{Young}
 & & & & & $\times$ \cr
 \cr
\end{Young}}}
\
&=& K([43]1[331]2|[61]1)\! \! \!
\raisebox{-9.0pt}{\mbox{
\begin{Young}
& & & $\times$\cr
& & \cr
\end{Young}}} \ \times \! \! \! \! \!
\raisebox{-15pt}{\mbox{
\begin{Young}
& & \cr
& & $\times$\cr
 \cr
\end{Young}}
}\nonumber \\ & &  + \ K([43]2[331]2|[61]1)\! \! \!
\raisebox{-9.0pt}{\mbox{
\begin{Young}
& & & \cr
& & $\times$ \cr
\end{Young}}} \ \times \! \! \! \! \!
\raisebox{-15pt}{\mbox{
\begin{Young}
& & \cr
& & $\times$\cr
 \cr
\end{Young}}}\nonumber \\
& & + \ K([43]2[331]3|[61]1)\! \! \!
\raisebox{-9.0pt}{\mbox{
\begin{Young}
& & & \cr
& & $\times$ \cr
\end{Young}}} \ \times \! \! \! \! \!
\raisebox{-15pt}{\mbox{
\begin{Young}
& & \cr
& & \cr
$\times$ \cr
\end{Young}} 
}\ ,
\end{eqnarray} 
where the cross in the left-hand side indicates that the 
states (\ref{singletch52}) and (\ref{prime})
correspond to $p = 2$ and  to $p = 1$ respectively.
We remind that in the right-hand side, if one removes the crossed box, 
the first and second Young tableaux 
describe the spin and flavor states respectively. In fact each such product
represents a spin-flavor state
of $S_6$ of partition $[6]$ and $[51]$  for 
$p$ = 2 and $p$ = 1  respectively, coupled to the 7th quark in a given
spin-flavor state. 
When $N_c = 3$ we recover the $^{2}1$ flavor singlet. This case is new and 
completes our work on the $[N_c-1,1]$ states by allowing to incorporate 
the $\Lambda$ baryons.

We recall that the approximate wave function \cite{CCGL} contains only terms 
with $p$ = 2 as discussed in Ref.\ \cite{Matagne:2006dj}.

The isoscalar factors $K([f']p'[f'']p''|[f]p)$ of $S_{N_c}$
for the spin-flavor states corresponding
$^{2}8$ and $^{4}8$ and $^{2}10$  multiplets, when $N_c = 3$, 
have been obtained in Ref. \cite{Matagne:2008fw} 
and as we need them again, for self-consistency they are  reproduced   
in Appendix A.

The analytic forms obtained here for the isoscalar factors 
needed for the states corresponding to the flavor singlet $^{2}1$  
when $N_c$ = 3 are reproduced in Table \ref{spinonehalf}.
Details of the calculations are given in Appendix B.
Note that the analytic expressions of Table \ref{spinonehalf} hold
for $N_c$ odd only, because the partitions must contain integer numbers.

In Table \ref{spinonehalf} and those of Appendix \ref{appa}, 
the isoscalar factors from the columns with $p = 1$ and $ p = 2$
obey the orthogonality property  defined generally as
\begin{eqnarray}\label{orthog}
\sum_{p'p''}  K([f']p'[f'']p''|[f]p) K([f']p'[f'']p''|[f_1]p_1) & 
= &\delta_{f f_1}
\delta_{p p_1}. \label{K1} 
\end{eqnarray}
The expressions exhibited in Table \ref{spinonehalf} have been  
checked for $N_c$ = 3, 5, 7 
and 9, by using the recurrence relation described in Ref. \cite{ISOSC} which 
allows to obtain isoscalar factors of $S_{N-1}$ from those of $S_{N}$. 
For consistency the same phase convention must be constantly applied.
Tables  \ref{spinonehalf}, \ref{spin1/2n}, \ref{spin3/2n} and \ref{spin1/2delta}
prove that this is the 
case, one has the same phase irrespective of $N_c$. Thus they offer a convenient 
test to check the phase convention rule. The results of Table \ref{spinonehalf}
and those reproduced in Appendix A can be used for any fermion system 
described by the partition $[N_c - 1, 1]$ where one fermion must
be separated from the rest. 

\begin{table}
\caption{Isoscalar factors $K([f']p'[f'']p''| [f]p)$ for
$S = 1/2$, corresponding to $^{2}1$ when $N_c=3$. }\label{spinonehalf} 
\renewcommand{\arraystretch}{1.8}
 \begin{tabular}{c|ccc}
$[f']p'[f'']p''$   & $[N_c-1,1]1$  &\hspace{3cm}& $[N_c-1,1]2$  \\ \hline
$\left[ \frac{N_c+1}{2},\frac{N_c-1}{2}\right ] 1\left\lbrack
\frac{N_c-1}{2},\frac{N_c-1}{2},1\right \rbrack 2$ 
& $-\frac{1}{2}\sqrt{\frac{(N_c-3)(N_c+1)}{N_c(N_c-2)}}$ &  & $ 0 $ \\
$\left[ \frac{N_c+1}{2},\frac{N_c-1}{2}\right ] 1\left\lbrack
\frac{N_c-1}{2},\frac{N_c-1}{2},1\right \rbrack 3$ 
& $ 0 $ &  & $ 1 $ \\
$\left\lbrack \frac{N_c+1}{2},\frac{N_c-1}{2}\right \rbrack 2 
\left\lbrack \frac{N_c-1}{2},\frac{N_c-1}{2},1\right \rbrack 2$ 
& $\frac{1}{2}\sqrt{\frac{3(N_c-3)(N_c+1)}{N_c(N_c-2)}}$  
&\hspace{1cm} & $ 0$ \\
$\left\lbrack \frac{N_c+1}{2},\frac{N_c-1}{2}\right \rbrack 2 
\left\lbrack \frac{N_c-1}{2},\frac{N_c-1}{2},1\right \rbrack 3$ 
&   $ -\sqrt{\frac{3}{N_c(N_c-2)}} $ & & 0 \\
\hline
\end{tabular}
\end{table}

\section{Matrix elements}\label{matelem}

It is very important to apply the $1/N_c$ expansion
method to both nonstrange and strange baryons together.
First, we have at our disposal a larger number of experimental
data than for nonstrange baryons alone 
and second, we can get an unified picture of all light baryons.
 
In the following we consider $N_f$ = 3. When the SU(3)-flavor symmetry is exact, the $1/N_c$ expansion 
mass operator describing an excited state 
can be written as the linear combination 
\begin{equation}
\label{massoperator1}
M^{(1)} = \sum_{i} c_i O_i ,
\end{equation} 
where $c_i$ are unknown coefficients which parametrize the QCD dynamics
and the operators $O_i$ are 
combinations of SU(6) and SO(3) generators $L^i$. The presence 
of $L^i$ is necessary in describing excited states.

For the purpose of our analysis  and as an extension of the 
previous work \cite{Matagne:2008fw}, here it is enough to consider some of 
the most dominant operators, namely the spin 
and flavor operators. Previous experience indicates that
the most dominant operators to order $\mathcal{O}({1/N_c})$ included, 
are those constructed from SU(2$N_f$) exclusively
\cite{Matagne:2006dj}, the operators containing   $L^i$ bringing 
usually  smaller contributions.

We recall that in the standard procedure, based on core+quark separation,
these operators 
are $s \cdot S_c$,~ $S^2_c$,~ $t \cdot T_c $ and $T^2_c$. 
The analytic expressions of the expectation values, calculated both with 
the approximate and the exact wave functions, as 
defined in the previous section, are presented in Tables \ref{spinop} 
and \ref{flavorop}.

Regarding the spin operators the only change with respect to
SU(4) \cite{Matagne:2008fw} is the addition of the last row
where the result is naturally identical to that of 
first and third ones when  the exact wave function is used, because
$^28$, $^210$ and $^21$ have the same spin. The approximate wave function
leads to different results for $^28$, $^210$ and $^21$ because the
wave function is truncated. Note that for the approximate
wave function we agree with Ref. \cite{Goity:2002pu}. As a matter of fact, 
for the approximate wave function, the matrix elements of $s \cdot S_c$
and $S^2_c$ are independent of $N_c$ for $^48$, $^210$ and $^21$
the reason being again the truncation of the wave function.

\begin{table}
{
 \renewcommand{\arraystretch}{1.5}
\caption{Matrix elements of the spin operators calculated with the 
approximate and the exact wave functions.}\label{spinop}
\begin{tabular}{c||c|c|c|c|} 
\hline  
 & \multicolumn{2}{c|}{$\langle s\cdot S_c\rangle$} & \multicolumn{2}{c|}{$\langle S^2_c\rangle$}\\ \cline{2-5}
& approx. w.f. & exact w.f. & approx. w.f. & exact w.f. \\ \hline
$^2 8$ & $ -\frac{N_c+3}{4N_c}$ & $-\frac{3(N_c-1)}{4N_c}$ & $\frac{N_c+3}{2N_c}$  & $\frac{3(N_c-1)}{2N_c}$ \\ 
$^4 8$ & $\frac{1}{2}$  & $-\frac{3(N_c-5)}{4N_c}$  & 2 & $\frac{3(3N_c-5)}{2N_c}$  \\ 
$^2 10$ &  $-1$ & $-\frac{3(N_c-1)}{4N_c}$& 2 &  $\frac{3(N_c-1)}{2N_c}$ \\ 
$^2 1 $ & $0$  & $-\frac{3(N_c-1)}{4N_c}$  & $0$ & $\frac{3(N_c-1)}{2N_c}$ \\ 
\hline \hline
\end{tabular}}
\end{table}
\begin{table}\label{flavorop}
\caption{Matrix elements of the flavor operators calculated with the 
approximate and the exact wave functions.}
\renewcommand{\arraystretch}{1.5}
\begin{tabular}{c||c|c|c|c|} \cline{2-5}
 & \multicolumn{2}{c|}{$\langle t\cdot T_c\rangle$} & \multicolumn{2}{c|}{$\langle T^2_c\rangle$}\\ \cline{2-5}
& approx. w.f. & exact w.f. & approx. w.f. & exact w.f. \\ \hline
$^2 8$ & $\frac{N_c(N_c-4)-9}{12N_c} $ & $\frac{(N_c-9)(N_c-1)}{12N_c} $ & $\frac{18+N_c+4N_c^2+N_c^3}{12N_c} $  & $ \frac{(N_c-1)[18+N_c(N_c+5)]}{12N_c}$ \\ 
$^4 8$ &  $\frac{N_c-13}{12} $ & $\frac{(N_c-9)(N_c-1)}{12N_c} $& $\frac{19+N_c(N_c+4)}{12} $ &  $\frac{(N_c-1)[18+N_c(N_c+5)]}{12N_c} $ \\ 
$^2 10$ & $\frac{N_c+5}{12} $  & $\frac{45+N_c(N_c-10)}{12N_c} $  & $\frac{(N_c+2)^2+15}{12} $ & $\frac{-90+N_c[49+N_c(N_c+4)]}{12N_c} $  \\
$^2 1$ & $-\frac{N_c+5}{6}$ & $\frac{N_c(N_c-16)-9}{12N_c}$ & $\frac{(N_c-1)(N_c+5)}{12}$ & $\frac{18+N_c[7+N_c(N_c-2)]}{12N_c}$ \\ 
 \hline \hline
\end{tabular}
\end{table}

The expressions and the order in $N_c$ of the expectation values
of $t \cdot T_c $ and $T^2_c$  with  SU(6)
wave functions  are naturally different
from those of SU(4) \cite{Matagne:2008fw}.
Using the wave function described by Eqs. (\ref{fs})-(\ref{flavor})
we have first obtained the general form of the expectation value of $T^2_c$ 
at fixed $p$. This is 
\begin{equation}
\langle T^2_c \rangle_p = \frac{1}{3}\sum_{p' p''} 
 K([f']p'[f'']p''|[f]p)^2 g_{\lambda_c\mu_c}
\end{equation}
where 
\begin{equation}\label{glm}
g_{\lambda_c\mu_c} = (\lambda^2_c + \mu^2_c + \lambda_c \mu_c 
+ 3 \lambda_c + 3 \mu_c),
\end{equation}
where $\lambda_c $  and $\mu_c  $ depend on $p''$ as defined below 
Eq. (\ref{flavor}).

Taking $p$ = 2 we recover the expressions of $\langle T^2_c \rangle$
with the  approximate wave function as exhibited in Table  \ref{flavorop},
column 3.

For the exact wave function both $p$ = 1 and $p$ = 2 contribute.
According to the discussion following Eq.  (\ref{EWF}) the
expectation value of  $T^2_c$ becomes
\begin{equation}
\langle T^2_c \rangle = \frac{1}{N_c - 1}
\left[\langle T^2 \rangle_{p=2} + (N_c - 2)  \langle T^2 \rangle_{p=1}\right].
\end{equation}
Note that such a combination of  $p$ = 1 and $p$ = 2 terms is required for
any operator in the mass formula (\ref{massoperator1}) when the
matrix elements are calculated with the exact wave function. For $\langle T^2_c \rangle$
the results are presented in the last column of Table \ref{flavorop}.

Knowing that $\langle T^2 \rangle = g_{\lambda \mu}/3$ 
with $g_{\lambda \mu}$ defined as in Eq. (\ref{glm}),
but with  $ {\lambda_c \mu_c} \rightarrow {\lambda \mu}$,
one can then derive the matrix element of $t \cdot T_c$ as
\begin{equation}
\langle t \cdot T_c \rangle = \frac{1}{2}
\left[\langle  T^2 \rangle - \langle T^2_c \rangle - \frac{4}{3}\right],
\end{equation}
both for the exact and the approximate wave functions.
At fixed $p$ this is in agreement with Eq. (A15) of 
Ref. \cite{Matagne:2006zf}.

Actually we are interested in the operators $\frac{1}{N_c} ~t \cdot T_c $
and  $\frac{1}{N_c} ~T^2_c$, entering the mass formula (\ref{massoperator1}).
One can see that with the exact 
wave functions the matrix elements
of the operator  $\frac{1}{N_c} ~t \cdot T_c $
are of  order $\mathcal{O}({N^0_c})$ and those of the operator  
$\frac{1}{N_c} ~T^2_c$
of order $\mathcal{O}({N_c})$  for all spin-flavor multiplets of mixed 
symmetric states. 
To fulfill the large 
$N_c$ counting  a solution would be to make the replacement
\begin{equation}\label{tTc} 
\frac{1}{N_c} t \cdot T_c \rightarrow \frac{1}{N_c} 
\left( t \cdot T_c - \frac{1}{12} \openone \right), 
\end{equation} 
where the shift is due to the subtraction of the dominant
operator $\mathcal{O}_1 =  \openone$ of order  $\mathcal{O}(N_c)$,
and similarly, $\frac{1}{N_c} ~T^2_c$ can be replaced by 
\begin{equation}\label{Tc2} 
\frac{1}{N_c} T^2_c  \rightarrow \frac{1}{N_c} 
\left( T^2_c - \frac{N_c}{12} \openone \right), 
\end{equation}
because for both operators the extracted terms are identical for all 
spin-flavor multiplets of the mixed representation $[N_c-1,1]$. 
By compensation, in the mass operator, these terms can provide 
an additional contribution to the leading orders $N^0_c$ and $N_c$ 
respectively.

A similar procedure is impossible for the approximate wave function
because the $^21$ multiplet has a different large $N_c$ analytic form than the
other spin-flavor multiplets of $[N_c-1,1]$, as one can see from Table \ref{flavorop}.
Thus there is no unique term to be subtracted.

Based on the standard procedure with the approximate 
wave function, the authors of Ref.  \cite{Cohen:2005ct}  observed that the 
matrix elements of the shifted operator (\ref{tTc}), denoted in their work
by $O_5$  vanish for ``all states
in multiplets with $Y_{max} = \frac{N_c}{3}$ (which includes all nonstrange
states in the ``${\bf 70}$"). For $Y_{max} = \frac{N_c}{3} - 1$ multiplets
the matrix elements of $O_5$ are found to be $-1/4$". 
The latter value is consistent with the expression in the last row 
of our Table \ref{flavorop}  for the approximate wave function because 
$\langle t \cdot T_c \rangle \rightarrow - N_c/6 $
in the large $N_c$ limit. Note that the vanishing of the expectation 
values of the operator  
(\ref{tTc}) takes place in fact only at large $N_c$  
for the $^28$, $^48$ and $^210$ multiplets.
This makes us to
believe  that,  to some extent, it was known that the $^21$ multiplet
had a different large $N_c$ behavior than the other
multiplets, but from the statement of Ref.  \cite{Cohen:2005ct}
it is not quite clear that the 
cancellation takes place at large $N_c$ only and that it does not hold 
for the $^21$ multiplet of ``${\bf 70}$".

Therefore from the analysis of Table \ref{flavorop} we conclude that the 
approximate wave function does not lead to same the large $N_c$ 
limit for 
$\langle t \cdot T_c \rangle$, irrespective of the spin-flavor multiplet 
contained in $[N_c - 1,1]$. In addition, note that for the $^21$ multiplet  
the sign of  the matrix element of $\langle t \cdot T_c \rangle $ is negative,  
consistent with Ref. \cite{Cohen:2005ct}.
However, let us note that  in the symmetric core approach
there is no problem in obtaining five towers of states 
because  $t \cdot T_c/N_c$ appears among the four $\mathcal{O}(N^0_c)$
needed operators.

We have already encountered the problem with the large $N_c$ behavior  
of $T^2$ in Ref. \cite{Matagne:2008kb}.
Based on the correspondence between a Young diagram and an irrep of 
SU(3), we have shown that
using SU(6) generators  acting on the whole system, 
here  $T^a$ of Eq. (\ref{generators}),
a general elegant solution is to redefine
the operator $T^aT^a$ as 
\begin{equation}\label{shift}
\frac{1}{N_c} \left[ T^aT^a-\frac{1}{12}N_c(N_c+6) \right]
\end{equation} 
where the subtracted term, $\frac{1}{12}N_c(N_c+6)$, appears in the 
Casimir expectation value of all irreducible representations 
of SU(3) contained in $\bf {70}$. The operator (\ref{shift}) is of order 
$N^0_c$ for $N_f$ = 3. Interestingly, this definition recovers the expectation 
values of the isospin operator 
$O_4 = \frac{1}{N_c} T^i T^i$ ($i = 1,2,3$) 
for  $N_f$ = 2 \cite{Matagne:2006dj}. Indeed, from the expectation value  
of the first term with 
$f$ = 0 in Eq. (30) of Ref. \cite{Matagne:2008kb} we obtain
\begin{equation}
\langle O_4 \rangle = \frac{1}{4 N_c} \lambda (\lambda + 2).
\end{equation}
Taking  $\lambda = 2 I$ we recover the SU(4) results
of Table 3, column 5 of Ref. \cite{Matagne:2006dj} in a single formula,
which shows that the matrix elements of the isospin operator are order $1/N_c$,
as expected. 

But, on the other hand, for the flavor singlet $^21$ the eigenvalues 
of the operator (\ref{shift}) are of order $\mathcal{O}(N^0_c)$.
This can be seen by writing the eigenvalue of the Casimir operator
not in terms of $\lambda$ and $\mu$ but in terms of $\lambda$
and $f$, where $f$ is the number of columns
filled with 3 boxes and $\mu=(N_c-\lambda-3f)/2$ \cite{Matagne:2008kb}.
In that case we have \cite{Matagne:2008kb}
\begin{equation}\label{ttsu3}
 \frac{1}{N_c}T^aT^a = 
 \frac{1}{12N_c}\left\{N_c(N_c+6)+3\lambda(\lambda+2)-3f[2(N_c+3)-3f]\right\}.
\end{equation}
From this expression the first term cancels out with the last term in
Eq. (\ref{shift}) and the remaining quantity is of order $\mathcal{O}(N^0_c)$.
Therefore the flavor operator introduces a shift between the 
flavor singlet $^21$ and the other multiplets, namely $^28$, $^48$
and $^210$.  This is entirely consistent with the results of Cohen and
Lebed \cite{Cohen:2005ct} of five irreps labelled by the
grand spin $K$.
Then the 5 independent mass eigenvalues are split by $\mathcal{O}(N^0_c)$.
By analogy, in our approach the five towers will be due to
the operator  $O_1 = N_c \ \1 $  of order $\mathcal{O}(N_c)$
and to 4 other operators of order 
$\mathcal{O}(N^0_c)$ which are $L \cdot S$, 
~$\frac{1}{N_c} \left[ T \cdot T-\frac{1}{12}N_c(N_c+6) \right]$,
~$\frac{1}{N_c} L \cdot T \cdot G$ and  
$\frac{1}{N_c} L^{(2)} \cdot G \cdot G $. 
Their matrix elements will be presented elsewhere.

Moreover, from Table \ref{flavorop} one can see that the matrix elements 
of $t \cdot T_c$ and $T^2_c$ with the approximate wave function
associated with the octets $^28$ and $^48$
are different from each other, which is not 
natural, because the flavor operators are independent of spin. By contrast,
the exact wave function leads to identical expressions, which is correct.
In this situation, a quantitative discussion between results 
obtained, on the one hand, with the exact, and on the other hand, with the 
approximate wave functions, cannot be made, as it was done for SU(4).

\section{Conclusions}

Let us briefly present our conclusions. In the scheme based on the  
decoupling of the system into a symmetric core of $N_c-1$ quarks
and an excited quark,
the flavor operator 
was decomposed in three independent parts, namely $T^2_c$, $t \cdot T_c$
and a constant both in SU(4) \cite{CCGL} and in SU(6) \cite{Goity:2002pu}.

In SU(4), describing  nonstrange baryons, $S^2_c$ and $T^2_c$ have identical 
matrix elements because
the spin and isospin states of a symmetric core are identical, so that
$T^2_c$ can be neglected. In SU(6) the situation
is different. One must include $T^2_c$ as well in the mass formula.

Thus in SU(4), in the decoupling scheme, the isospin can be  introduced
only through 
the operator $t \cdot T_c$. However  $t \cdot T_c$
has been ignored in all studies of the spectrum based on 
this  scheme, although it has entirely different matrix
elements from those of $s \cdot S_c$ \cite{CCGL}.
In Ref. \cite{Matagne:2008fw}, Table VI,
we have shown that the introduction of  
$\frac{1}{N_c}t \cdot T_c$, together with $\frac{1}{N_c}S^2_c$ 
and $\frac{1}{N_c}s \cdot S_c$ as independent operators, deteriorates the fit.
This may explain why $\frac{1}{N_c}t \cdot T_c$ has been avoided in previous 
numerical fits  in SU(4) \cite{CCGL}.

In SU(6) both $t \cdot T_c$ and $ T^2_c $ are necessary and 
both have been ignored \cite{Goity:2002pu}. Actually it 
would be simpler and more physically to consider the full Casimir operator  
$T^2$ of SU(3) instead of independent $t \cdot T_c$ and $ T^2_c $ operators.
Such a procedure was used in 
SU(4) \cite{Matagne:2006dj} for the isospin operator .

Due to the failure of the approximate wave functions \cite{CCGL} to lead to
the same large $N_c$ counting of $\frac{1}{N_c}t \cdot T_c$ 
for all spin-flavor multiplets of ${\bf 70}$,
a quantitative estimate between 
the two approaches, one based on the exact, the other on the 
approximate wave function, could not presently be made for SU(6) as it was 
previously made for SU(4) \cite{Matagne:2008fw}.

\vspace{1cm}
\noindent
{\bf Acknowledgments}.
N. M. acknowledges financial support from F.R.S. - FNRS (Belgium).

\appendix

\section{Isoscalar factors needed for 
$^{2}8$, $^{4}8$ and $^{2}10$}\label{appa}

Here we find it useful to recall the isoscalar factors of $S_N$ obtained in 
Ref. \cite{Matagne:2008fw} which are also needed in this study.
They are exhibited in Tables \ref{spin1/2n}, \ref{spin3/2n}
and \ref{spin1/2delta}
for $S=I=1/2$, $S=3/2,\ I=1/2$ and $S=1/2,\ I=3/2$ states respectively,
for arbitrary $N_c$.

\begin{table}
\caption{Isoscalar factors $K([f']p'[f'']p''| [f]p)$ for
$S=I=1/2$, corresponding to $^{2}8$ when $N_c=3$. 
}\label{spin1/2n}
\renewcommand{\arraystretch}{1.8}
 \begin{tabular}{c|ccc}
$[f']p'[f'']p''$   & $[N_c-1,1]1$  &\hspace{3cm}& $[N_c-1,1]2$  \\ \hline
$\left[ \frac{N_c+1}{2},\frac{N_c-1}{2}\right ] 1\left\lbrack \frac{N_c+1}{2},\frac{N_c-1}{2}\right \rbrack 1$ &    0    &  & $-\sqrt{\frac{3(N_c-1)}{4N_c}}$ \\
$\left\lbrack \frac{N_c+1}{2},\frac{N_c-1}{2}\right \rbrack 2 \left\lbrack \frac{N_c+1}{2},\frac{N_c-1}{2}\right \rbrack 2$ & $\sqrt{\frac{N_c-3}{2(N_c-2)}}$ &\hspace{1cm} & $\sqrt{\frac{N_c+3}{4N_c}}$ \\
$\left\lbrack \frac{N_c+1}{2},\frac{N_c-1}{2}\right \rbrack 2 \left\lbrack \frac{N_c+1}{2},\frac{N_c-1}{2}\right \rbrack 1$ &   $-\frac{1}{2}\sqrt{\frac{N_c-1}{N_c-2}}$ & & 0 \\
$\left\lbrack \frac{N_c+1}{2},\frac{N_c-1}{2}\right \rbrack 1 \left\lbrack \frac{N_c+1}{2},\frac{N_c-1}{2}\right \rbrack 2$ &   $-\frac{1}{2}\sqrt{\frac{N_c-1}{N_c-2}}$ & & 0 \\
\hline
 \end{tabular}
\end{table}

\begin{table}
\caption{Isoscalar factors $K([f']p'[f'']p''| [f]p)$ for
$ S=3/2,\ I=1/2$, corresponding to $^{4}8$ when $N_c=3$. 
}\label{spin3/2n}
\renewcommand{\arraystretch}{1.8}
 \begin{tabular}{c|ccc}
$[f']p'[f'']p''$   & $[N_c-1,1]1$  &\hspace{3cm}& $[N_c-1,1]2$  \\ \hline
$\left\lbrack \frac{N_c+3}{2},\frac{N_c-3}{2}\right \rbrack 1 \left\lbrack \frac{N_c+1}{2},\frac{N_c-1}{2}\right \rbrack 1$ & $\frac{1}{2}\sqrt{\frac{(N_c-1)(N_c+3)}{N_c(N_c-2)}}$ & \hspace{1cm} & 0 \\
$\left\lbrack \frac{N_c+3}{2},\frac{N_c-3}{2}\right \rbrack 2 \left\lbrack \frac{N_c+1}{2},\frac{N_c-1}{2}\right \rbrack 2$ & $\frac{1}{2}\sqrt{\frac{5(N_c-1)(N_c-3)}{2N_c(N_c-2)}}$ & \hspace{1cm} & 0 \\
$\left\lbrack \frac{N_c+3}{2},\frac{N_c-3}{2}\right \rbrack 1 \left\lbrack \frac{N_c+1}{2},\frac{N_c-1}{2}\right \rbrack 2$ & $\frac{1}{2}\sqrt{\frac{(N_c-3)(N_c+3)}{2N_c(N_c-2)}}$ & & 1 \\
 $\left\lbrack \frac{N_c+3}{2},\frac{N_c-3}{2}\right \rbrack 2 \left\lbrack \frac{N_c+1}{2},\frac{N_c-1}{2}\right \rbrack 1$ & 0 & & 0 \\
\hline
 \end{tabular}
\end{table}

\begin{table}
\caption{Isoscalar factors $K([f']p'[f'']p''| [f]p)$ for
$S=1/2,\ I=3/2$, corresponding to $^{2}10$ when $N_c=3$.
}\label{spin1/2delta}
\renewcommand{\arraystretch}{1.8}
 \begin{tabular}{c|ccc}
$[f']p'[f'']p''$   & $[N_c-1,1]1$  &\hspace{3cm}& $[N_c-1,1]2$  \\ \hline
$\left\lbrack \frac{N_c+1}{2},\frac{N_c-1}{2}\right \rbrack 1 \left\lbrack \frac{N_c+3}{2},\frac{N_c-3}{2}\right
\rbrack 1$& $\frac{1}{2}\sqrt{\frac{(N_c-1)(N_c+3)}{N_c(N_c-2)}}$ & \hspace{1cm} & 0 \\
$\left\lbrack \frac{N_c+1}{2},\frac{N_c-1}{2}\right \rbrack 2 \left\lbrack \frac{N_c+3}{2},\frac{N_c-3}{2}\right \rbrack 2$ & $\frac{1}{2}\sqrt{\frac{5(N_c-1)(N_c-3)}{2N_c(N_c-2)}}$ & \hspace{1cm} & 0 \\
$\left\lbrack \frac{N_c+1}{2},\frac{N_c-1}{2}\right \rbrack 2 \left\lbrack \frac{N_c+3}{2},\frac{N_c-3}{2}\right \rbrack 1$ & $\frac{1}{2}\sqrt{\frac{(N_c-3)(N_c+3)}{2N_c(N_c-2)}}$ & & 1 \\
$\left\lbrack \frac{N_c+1}{2},\frac{N_c-1}{2}\right \rbrack 1 \left\lbrack
\frac{N_c+3}{2},\frac{N_c-3}{2}\right \rbrack 2$ & 0 & & 0 \\
\hline
\end{tabular}
\end{table}


\section{Equations for isoscalar factors needed for $^{2}1$}
Here we shortly describe the method used to obtain general
analytic forms for the isoscalar factors of $S_{N_c}$ 
presented in Table \ref{spinonehalf}. 

Recall that we deal with  the partitions
\begin{equation}\label{partition}
[f'] = \left[\frac{N_c+1}{2},\frac{N_c-1}{2}\right],
~~[f'']= \left[\frac{N_c-1}{2},\frac{N_c-1}{2},1\right],
~~[f] =  \left[N_c-1,1\right].
\end{equation}

The case $p = 2$ is trivial for $^21$, $^48$ and $^210$, see Tables 
\ref{spinonehalf},    \ref{spin3/2n}   and  \ref{spin1/2delta}.
Using Young diagrams it becomes obvious that there is only one 
non-vanishing isoscalar factor, which according to the orthogonality relation
(\ref{orthog}) reads 
\begin{equation}\label{p2}
K([f']1[f'']3|[f]2) = 1.
\end{equation}
For example, Eq. (\ref{singletch52}) contains a single term
in the right-hand side and therefore the corresponding isoscalar factor 
has the value 1, consistent with the normalization properties and the phase
convention \cite{Stancu:1991rc,ISOSC}.

For $p = 1$ there are three non-vanishing isoscalar factors, so we need 
three equations to derive them. Our first idea was to use the same equations 
as in Ref. \cite{Matagne:2008fw}. They turn out to be quadratic. 

The first equation is the orthogonality relation (\ref{orthog})
which for  $p = 1$ becomes   
\begin{equation}\label{norm}
(K([f']1[f'']2|[N_c-1,1]1))^2 + (K([f']2[f'']2|[N_c-1,1]1))^2 
+ (K([f']2[f'']3|[N_c-1,1]1))^2 = 1.
\end{equation}

A second equation is provided by the diagonal matrix elements of the SU(6) 
generator $G^{ia}$, written as a sum of two generators: 
$G^{ia}_c$ which acts on a subsystem of $N_c-1$ quarks, where 
$c$ stands for "core", and  $g^{ia}$ which acts on the last quark 
separated from the rest. One has
\begin{equation}
 \langle G^{ia} \rangle = \langle G^{ia}_c \rangle + \langle g^{ia} \rangle
\end{equation}
Note that for $p = 1$ the core has the
partition  $[N_c-2,1]$. 
This equation gives the following relation between the isoscalar factors
\begin{eqnarray}\label{split}
\lefteqn{ \frac{N_c+3}{4}(K([f']1[f'']2|[N_c-1,1]1))^2 -\frac{N_c(N_c-6)-11}{4(N_c-1)}(K([f']2[f'']2|[N_c-1,1]1))^2}  \nonumber \\ & + & \frac{(N_c-3)(N_c+3)}{2(N_c-1)}(K([f']2[f'']3|[N_c-1,1]1))^2\nonumber \\  &+ &\frac{\sqrt{3}(N_c-1)}{2}K([f']1[f'']2|[N_c-1,1]1)K([f']2[f'']2|[N_c-1,1]1) \nonumber \\
& + & \frac{4\sqrt{(N_c-3)(N_c+1)}}{N_c-1}K([f']2[f'']2|[N_c-1,1]1)K([f']2[f'']3|[N_c-1,1]1)\nonumber \\ &+ & \frac{N_c-3}{2}=0
\end{eqnarray}
From the Casimir identity, one can derive a third equation
\begin{equation}\label{casimir}
 \frac{\langle s\cdot S_c\rangle}{3} + \frac{\langle t\cdot T_c\rangle}{2}+2\langle g\cdot G_c\rangle = \frac{5N_c-11}{12}, 
\end{equation}
which gives another quadratic equation 
\begin{eqnarray}\label{eq3}
\lefteqn{\frac{(N_c-1)^2}{48}(K([f']1[f'']2|[N_c-1,1]1))^2} \nonumber \\ & + &\frac{83+N_c(N_c(N_c-15)+7)}{48(N_c-1)}(K([f']2[f'']2|[N_c-1,1]1))^2\nonumber \\    &- &  \frac{91-N_c(N_c(N_c+3)-9)}{48(N_c-1)}(K([f']2[f'']3|[N_c-1,1]1))^2 \nonumber \\ & + & \frac{\sqrt{3}(N_c-1)}{4} K([f']1[f'']2|[N_c-1,1]1)K([f']2[f'']2|[N_c-1,1]1) \nonumber \\ & + &
\frac{2\sqrt{(N_c-3)(N_c+1)}}{N_c-1}K([f']2[f'']2|[N_c-1,1]1)K([f']2[f'']3|[N_c-1,1]1)\nonumber \\
& - & \frac{(N_c-19)(N_c-1)}{48} = 0.
\end{eqnarray}

However, one can show that the system of equations (\ref{norm}), (\ref{split}) and (\ref{casimir}) is linearly dependent. By using a recurrence relation described in Refs. 
\cite{ISOSC,Stancu:1991rc}, we have derived the values of the isoscalar factors 
for $N_c = 3, 5, 7$ and $9$. From them we could quite easily make a 
generalization to an arbitrary $N_c$. These expressions  
are presented in Table \ref{spinonehalf}. They verify the three equations 
presented above.

\section{Matrix elements of $\langle g \cdot G_c \rangle$}
Equation (\ref{eq3}) requires the knowledge of the matrix element 
of $\langle g \cdot G_c \rangle $ for which we first need the matrix 
elements of $G_c^{ja}$. Using the generalized Wigner-Eckart theorem,
the matrix elements of $G_c^{ja}$ have been obtained in
Ref. \cite{Matagne:2008kb} as 
\noindent
\begin{eqnarray}\label{CORE}
 \lefteqn{\langle [N_c-1,1]p;(\lambda'\mu')Y'I'I_3';S'm_s'|G_c^{ja}|[N_c-1,1]p;(\lambda\mu)YII_3;Sm_s\rangle =}\nonumber\\
&  & (-1)^{1/2-S}\sqrt{(2S+1)}\sqrt{\frac{C^{[f_c]}(SU(6))}{2}}
\left(\begin{array}{cc|c}
 S & 1& S'\\
m_s & j & m'_s
\end{array}\right)
\left(\begin{array}{cc|c}
 I & I^a& I'\\
I_3 & I^a_3 & I'_3
\end{array}\right) \nonumber \\
& &\times \sum_{p',p'',p'_1,p^{''}_1} (-1)^{S'_c}(-1)^{\lambda-\lambda_c+\lambda'-\lambda'_c}(-1)^{\mu-\mu_c+\mu'-\mu'_c}\sqrt{(2S_c'+1)}K([f']p'[f'']p''|[N_c-1,1]p)\nonumber \\ 
& &\times K([f']p'_1[f'']p^{''}_1|[N_c-1,1]p) 
\left\{\begin{array}{ccc}
       S & 1 & S' \\
	S'_c & 1/2 & S_c
      \end{array}\right\}
 \sum_{\rho,\rho_c=1,2} \langle (\lambda\mu)YI;(11)Y^aI^a||(\lambda'\mu')Y'I'\rangle_\rho \nonumber \\
&  & \times U((10)(\lambda_c\mu_c)(\lambda'\mu')(11);(\lambda\mu)_\rho(\lambda'_c\mu'_c)_{\rho_c})
\left(\begin{array}{cc||c}
       [f_c] & [21^4] & [f_c] \\
	(\lambda_c\mu_c)S_c & (11)1 & (\lambda'_c\mu'_c)S'_c
      \end{array}\right)_{\rho_c} .
\end{eqnarray}
where $C^{[f_c]}(\mathrm{SU(6)}) $ is the SU(6)
Casimir operator associated to the irreducible representation 
described by the partition $[f_c]$. For $p =$ 1, 2 one has 
$[f_c] = [N_c-2,1]$ and  $[f_c] = [N_c-1]$ respectively, in agreement 
with the discussion in Sec. II. Note that in Ref. \cite{Matagne:2008kb}
there is a misprint which has been corrected 
here:  the factor $\sqrt{2S_c'+1}$ has been included in the sum over $p'$
in agreement with the definition of $S'_c$ following Eq. (\ref{fs}).
Also the upper index of the SU(6) Casimir operator has been corrected
by replacing  $[f]$ by $[f_c]$.

\begin{sidewaystable}
{\scriptsize
 \renewcommand{\arraystretch}{2.5}
\caption{Isoscalar factors of the SU(6) generator $G^{ja}$
needed for the $^21$ multiplet.} 
\label{isoscsu6}
\begin{tabular}{l|c|c|l}
\hline
\hline
$(\lambda_1\mu_1)S_1$ \hspace{0.5cm} & \hspace{0.5cm}$(\lambda_2\mu_2)S_2$ \hspace{0.5cm} & \hspace{0.5cm}$\rho$\hspace{0.5cm} & \hspace{0.5cm}$\left(\begin{array}{cc||c}                                         [N_c-1,1]  &  [21^4]  &  [N_c-1,1] \\
 (\lambda_1\mu_1)S_1 & (\lambda_2\mu_2)S_2 & (\lambda+1,\mu-2)S
                                      \end{array}\right)_\rho$  \\
\vspace{-0.8cm} &  &   & \\
\hline
$(\lambda+1,\mu-2)S$\hspace{0.5cm} & \hspace{0.cm}$(11)1$ & $1$ &\hspace{0.5cm} $\left[N_c(4S+7)+6(S+1)\right]\sqrt{\frac{2S}{(S+1)(N_c^2+12S(S+2))N_c(5N_c+18)}}$\\
$(\lambda+1,\mu-2)S$\hspace{0.5cm} & \hspace{0.cm}$(11)1$ & $2$ &\hspace{0.5cm}
$\frac{4(S(S+3)-1)-N_c(N_c+6)}{2S+1}\sqrt{\frac{3S(2S+1)(2S+3)(N_c+2S)(N_c-2(S+1))}{2(S+1)(N_c-2s)(N_c+2(S+2))(N_c^2+12S(S+2))N_c(5N_c+18)}}$\\
$(\lambda+1,\mu-2)S+1$\hspace{0.5cm} & \hspace{0.cm}$(11)1$ & $1$ &\hspace{0.5cm}
$3N_c\sqrt{\frac{2(2S+3)}{(S+1)(N_c^2+12S(S+2))N_c(5N_c+18)}}$ \\
$(\lambda+1,\mu-2)S+1$\hspace{0.5cm} & \hspace{0.cm}$(11)1$ & $2$ &\hspace{0.5cm}
$\left[12S(S+2)-N_c(N_c+6)\right]\sqrt{\frac{3(N_c+2S)(N_c-2(S+2))}{2(S+1)(2S+1)(N_c+2(S+2))(N_c^2+12S(S+2))(N_c-2S)N_c(5N_c+18)}} $\\
$(\lambda\mu)S+1$\hspace{0.5cm} & \hspace{0.cm}$(11)1$ & $/$ &\hspace{0.5cm}
$-\frac{S}{S+1}\sqrt{\frac{6(N_c-2(S+1))(2S+3)}{(2S+1)(N_c-2S)N_c(5N_c+18)}}$\\
\hline
\hline
\end{tabular}}
\end{sidewaystable}


For a single quark the matrix element of the generator $g^{ja}$
takes a simpler form \cite{Matagne:2008kb}
\begin{eqnarray}\label{SEPARATE}
 \lefteqn{\langle [N_c-1,1]p;(\lambda'\mu')Y'I'I_3';S'm_s'|g^{ja}|[N_c-1,1]p;(\lambda\mu)YII_3;Sm_s\rangle =}\nonumber\\
& &(-1)^{S'-1/2}\sqrt{2(2S+1)}\left(\begin{array}{cc|c}
 S & 1& S'\\
m_s & j & m'_s
\end{array}\right)
\left(\begin{array}{cc|c}
 I & I^a& I'\\
I_3 & I^a_3 & I'_3
\end{array}\right) \nonumber \\ &  & \times 
\sum_{p',p'',p'_1,p^{''}_1} (-1)^{S_c}K([f']p'[f'']p''|[N_c-1,1]p) 
K([f']p'_1[f'']p^{''}_1|[N_c-1,1]p)
\left\{\begin{array}{ccc}
       S & 1 & S' \\
	1/2 & S_c & 1/2
      \end{array}\right\} \nonumber \\
& & \times\sum_{\rho=1,2} 
\langle (\lambda\mu)YI;(11)Y^aI^a||(\lambda'\mu')Y'I'\rangle_\rho  
U((\lambda_c\mu_c)(10)(\lambda'\mu')(11);(\lambda\mu)_\rho(10)) .        
\end{eqnarray}
Note that in Ref. \cite{Matagne:2008kb} there is a misprint in the Racah
coefficient 
$U((\lambda_c\mu_c)(10)(\lambda'\mu')(11);(\lambda\mu)_\rho(10))$
where the lower index $\rho$ has been inadvertently shifted.

The above matrix elements lead to 
\begin{eqnarray}\label{ggc}
 \lefteqn{\langle [N_c-1,1]p;(\lambda'\mu')Y'I'I_3';S'm_s'|g\cdot G_c|
 [N_c-1,1]p;(\lambda\mu)YII_3;Sm_s\rangle = }\nonumber\\ & &
\delta_{SS'}\delta_{m_sm_s'} \delta_{II'}\delta_{I_3I_3'} \delta_{YY'} (-1)^{S-1/2}\sqrt{\frac{C^{[f_c]}(SU(6))}{2}}\nonumber\\ & &
\times \sum_{p',p'',p'_1,p^{''}_1} (-1)^{S_c}K([f']p'[f'']p''|[N_c-1,1]p)  K([f']p'_1[f'']p^{''}_1|[N_c-1,1]p)\nonumber \\ & & \times\sqrt{2(2 S_c'+ 1)}
\left\{\begin{array}{ccc}
       1/2& S_c' & S \\
	S_c & 1/2 & 1
      \end{array}\right\} 
 \nonumber \\ & & \times \sum_{\rho_c}
U((\lambda_c\mu_c)(11)(\lambda\mu)(10);(\lambda'_c\mu'_c)_{\rho_c}(10))
\left(\begin{array}{cc||c}
       [f_c] & [21^4] & [f_c] \\
	(\lambda_c\mu_c)S_c & (11)1 & (\lambda'_c\mu'_c)S'_c
      \end{array}\right)_{\rho_c}.
\end{eqnarray}
Eqs. (\ref{CORE}) or (\ref{ggc}) contain new isoscalar factors of SU(6)
which we derived in this study using the same method as in Ref.
\cite{Matagne:2008kb}. They  
are presented in Table  \ref{isoscsu6}
for $\rho$ = 1 and 2. The last row corresponds to $\rho$ = 1, the only
non-vanishing case. In a similar way we have also obtained 

\begin{equation}\label{new}
\left(\begin{array}{cc||c}
       [N_c-1,1] & [21^4] & [N_c-1,1] \\
	(\lambda-1,\mu+1)S & (11)1 & (\lambda-1,\mu-1)S
      \end{array}\right) =
 \frac{1}{S}\sqrt{\frac{6(2S-1)(S+1)(N_c-2(S-2))}{(N_c-2(S-1))N_c(5N_c+18)}}~,
\end{equation}
\noindent \\
where $\rho$, when unspecified, by convention 
corresponds to $\rho = 1$. When applying the above formulas one has to 
be rather careful with the
meaning of $\lambda$ and $\mu$ when they are expressed in terms of $S$
as above. 
For example for the flavor singlet $^21$ one has $S = 1/2$,
which implies that the label $(\lambda - 1, \mu - 1)$ corresponds to the 
flavor singlet, inasmuch as,  by definition, 
we take $\lambda = 2 S$ and $\mu = (N_c-2S)/2$. 
The isoscalar factors of Table \ref{isoscsu6} can be used in other
studies based on SU(6).


\end{document}